\documentclass[aps,prb,twocolumn,superscriptaddress]{revtex4}
\usepackage{graphicx}
\begin{document}
\title{Minigap in a SN junction with paramagnetic impurities}

\author{B. Crouzy}
\affiliation{Ecole Polytechnique Federale de Lausanne (EPFL), Institute for Theoretical Physics, CH-1015 Lausanne, Switzerland}
\author{E. Bascones}
\affiliation{Instituto de Ciencia de Materiales de Madrid, CSIC, Cantoblanco, E-28049 Madrid, Spain \\ Theoretische Physik, ETH-H\"onggerberg, CH-8093 Z\"urich, Switzerland}
\author{D. A. Ivanov}
\affiliation{Ecole Polytechnique Federale de Lausanne (EPFL), Institute for Theoretical Physics, CH-1015 Lausanne, Switzerland}

\date{\today}

\begin{abstract}
We study the effect of spin-flip scattering on the density of states 
in a long diffusive S-N 
junction with transparent interface. We calculate the critical value of 
the spin-flip scattering rate at which the minigap closes and give the
dependence of the minigap on the spin-flip rate. For the system we consider, the minigap and the critical spin-flip rate have the scale of the Thouless energy, and not of the superconducting gap.
\end{abstract}

\maketitle
Proximity structures made of superconducting and non-superconducting elements 
in contact with each other is a topic of current theoretical and experimental 
interest. Most of the experimental applications concern diffusive materials 
for which an appropriate formalism has been derived.\cite{Us} 
In the diffusive 
regime the motion of the electrons is governed by frequent 
scattering on impurity atoms: the elastic mean free path $l_{e}$ is much 
smaller than the relevant length scales of the system, namely the 
superconducting coherence length $\xi$ and the dimension of the normal (N) and 
superconducting (S) electrodes. 
As a consequence of Andreev reflection and its interplay with disorder, a gap 
is opened in the spectrum of the N-region (see for example \cite{Gb,Scheer,Gabino,Si} and references therein). This minigap does not depend on the position in the junction and 
is of the order of magnitude of the Thouless energy
$E_{\rm{Th}}=\frac{{D}}{L^2}$, where $D=\frac{v_{F}l_{e}}{3}$ is the 
diffusion constant and $L$ is the length of the normal wire. Here and in the following, we use the units with $\hbar=1$.

Elastic scattering on impurities is not harmful for isotropic
superconductivity and is directly connected to the appearance of the minigap.
On the contrary, scattering on magnetic impurities leads to the
destruction of singlet superconductivity.
The exchange interaction between the spin of the electrons and the magnetic 
impurities breaks the time reversal symmetry between the electrons in Cooper 
pairs and reduces the superconducting correlations.
Scattering on magnetic impurities can be described in terms of a spin-flip 
scattering rate $\Gamma_{\rm{sf}}$. In the Born approximation\cite{Ab1,Maki} 
$\Gamma_{\rm{sf}}$ is proportional to the impurity concentration. 
The energy gap $E_{g}$ in the electron density of states (DoS) of a bulk superconductor is lowered by 
spin-flip scattering 
\begin{equation}\label{ho1q} 
E_{g}=\Delta\left[1-\left(\frac{2\Gamma_{\rm{sf}}}{\Delta}\right)^{2/3}\right]^{3/2}
\end{equation}
and closes for the critical concentration of magnetic 
impurities\cite{Go2} $\Gamma_{\rm{sf}}^{\rm{bulk}}=\frac{\Delta}{2}$. Here, the order parameter $\Delta$ itself depends on $\Gamma_{\rm{sf}}$. 

In the ``dirty'' limit, the equations of motion for the quasiclassical matrix 
Green function (averaged over the fast Fermi oscillations and the momentum 
directions) can be reduced to the Usadel equations.\cite{Us,Belzig}
Usadel equations have been very useful in the analysis of position dependent 
problems like proximity effect.
Spin-flip and inelastic scattering processes can be included in the Usadel equations through the 
corresponding scattering rates $\Gamma_{\rm{sf}}$, respectively 
$\Gamma_{\rm{in}}$.\cite{Ko}

The impact of spin-flip scattering on the proximity effect and the minigap can 
be controlled by selectively doping the normal and/or superconducting regions 
with magnetic impurities or applying a magnetic field to a thin film or wire.\cite{Maki,reviewnano} 
Belzig {\it et al}\cite{Be} have numerically shown
 that the minigap is reduced and vanishes for large  values of $\Gamma_{\rm{sf}}$. For the case of a junction with $L=1.1\,\xi$ (corresponding to the Thouless energy of the order of $\Delta$), they find that the minigap closes when $\Gamma_{\rm{sf}} \approx 0.4\,\Delta$. Other works\cite{Yip,Yokoyama} have analyzed the effect of magnetic impurities in the transport properties of S-N junctions, finding that the Thouless energy is the scale which controls the effect of spin flip scattering on the proximity effect.
Due to the self-consistent treatment and the non-linearity of the Usadel 
equations the complete numerical solution of the problem is difficult. It is 
important to obtain simple expressions which allow comparison with experiment 
without solving the full equations for each set of parameters.

In this note we obtain such expressions for the case of a long junction. 
We show that a finite, but small $\Gamma_{\rm{sf}}$ suppresses linearly the 
minigap which closes at $\Gamma_{\rm{sf}}^{c}\approx0.62\,E_{\rm{Th}}$. We also 
present an asymptotic expression for the dependence of the  minigap on the 
spin-flip rate 
near the critical value. Above this value, we calculate the zero energy density of states.

We consider a finite size normal metal N, of length $L$ connected to a 
semi-infinite superconducting terminal S by a transparent interface. We assume 
that electronic motion is diffusive in both the normal and superconducting 
parts. We restrict our discussion to a quasi one-dimensional geometry/wire or to a thin film
and neglect any dependence on transversal coordinates.  
The origin of the coordinate $x$ is fixed at the S-N interface.
Introducing the angular parametrization\cite{Gb} $g^{R}=\cos\theta$, $f^{R}=\sin\theta$ 
for the normal, respectively the anomalous component of the retarded Green 
function, the Usadel equations in the N-region, where the pairing interaction 
vanishes, become
\begin{equation}\label{usad} 
\frac{D\partial_{x}^2\theta}{2}+(i\epsilon-\Gamma_{\rm{in}})\sin\theta-2\Gamma_{\rm{sf}}\cos\theta\sin\theta=0.
\end{equation}
The proximity angle $\theta$ is a function of the energy $\epsilon$ and the 
position $x$.  The electron DoS, in units of the normal state bulk value 
$\nu_{0}=\frac{mp_{f}}{2\pi^{2}}$, is given by
\begin{equation}\label{dd}
\frac{\nu(\epsilon,x)}{\nu_{0}}=\textrm{Re}\left[\cos\theta(\epsilon,x)\right].
\end{equation}
The boundary conditions for the quasiclassical equations have been discussed in Refs.~\onlinecite{Lu,Za}. At the interface with vacuum, the conservation of the quasiparticle current yields
\begin{equation}\label{bc2}
\partial_{x}\theta(x=L)=0.
\end{equation}

We will study these equations analytically using simplified boundary 
conditions at the  S-N interface, where we impose the superconducting bulk 
value (at zero energy) of the proximity angle
\begin{equation}\label{bc1}
\theta(x=0)=\frac{\pi}{2}.
\end{equation}
This boundary condition is justified for energies much smaller than the 
superconducting order parameter $\Delta$ and if the normal part is much more 
disordered than the superconducting part.\cite{Si}

Since the scale of the superconducting order parameter $\Delta$ does not 
appear in the rigid boundary condition (\ref{bc1}), we will write 
in the following the energies and the length in units of the only other 
relevant scale for our system: the Thouless energy $E_{\rm{Th}}$, respectively 
the width of the N-region $L$. 

The boundary conditions (\ref{bc2},\ref{bc1}) and the calculations presented here for the S-N 
junction can also be applied to describe S-N-S junctions 
with no phase difference between the superconducting terminals.\cite{Co2}
It is important to pay attention to the choice of the energy scale $E_{\rm{Th}}$. A S-N-S junction of length unity is equivalent to a S-N junction of length $\frac{1}{2}$. Therefore all the energies must be multiplied by a factor four if we consider S-N-S junctions.\\

In the limit of small energies and scattering rates the Usadel equation (\ref{usad}) can be linearized in terms of the  deviation 
$\delta\theta$ from the value of the proximity angle at the S-N interface (\ref{bc1}). At this level of approximation the 
DoS in the N-region is given by 
\begin{equation}\label{lin}
\frac{\nu(\epsilon,x)}{\nu_{0}}\approx\Gamma_{\rm{in}}^{\rm{eff}}\,x\,(2-x),
\end{equation}
where $\Gamma_{\rm{in}}^{\rm{eff}}=\Gamma_{\rm{in}}(1+2\Gamma_{\rm{sf}})$ and 
$x\in[0,1]$. 
We see that if $\Gamma_{\rm{in}}=0$ a small spin-flip scattering does not create states at small energies. If, on the contrary, $\Gamma_{\rm{in}}$ is finite, $\Gamma_{\rm{sf}}$ modifies the density of states (constant at small energies) induced by the inelastic scattering.

To study the presence of solutions where the electronic DoS (\ref{dd}) 
vanishes we focus in the rest of the paper on the situation with 
$\Gamma_{\rm{in}}=0$. We 
introduce a new notation for the proximity angle $\theta$ below the minigap 
and write $\theta=\frac{\pi}{2}+i\beta$ with $\beta$ real. The minigap $E_{g}$ 
is, by definition, the maximal energy compatible with a real $\beta$ and can 
be obtained\cite{Co2,Di} using a first integral of (\ref{usad})
\begin{equation}\label{first}
\partial_{x}\beta=2\sqrt{f(\beta^{\,1})-f(\beta)},
\end{equation}
where the superscript $1$ denotes the value at $x=1$ and $f(\beta)=\epsilon\sinh\beta+\Gamma_{\rm{sf}}\sinh^2\beta$. Integrating equation (\ref{first}) over the junction, we get 
\begin{equation}
\int_{0}^{\beta^{\,1}}\frac{d\beta}{2\sqrt{f(\beta^{\,1})-f(\beta)}}=1.\label{inte}
\end{equation}
Without spin-flip, we recover with this relation the well-known value\cite{Co2} of the minigap $E_{g}^0\approx 0.78$. The critical value of the spin-flip at which the minigap in the DoS closes is (see the discussion in Appendix)
\begin{equation}\label{ss}
\Gamma_{\rm{sf}}^{c}=\frac{\pi^2}{16}\approx0.62.
\end{equation}

For energies $\epsilon<E_{g}^0$ and $\Gamma_{\rm{sf}}=0$ equation (\ref{inte}) is solved for two 
different values of $\beta^{\,1}$. One of these solutions leads to a diverging 
$\beta^{\,1}$ when the energy goes to zero and we reject it using a continuity 
argument. This continuity argument is commonly accepted in the quasiclassical 
approximation, but the diverging branch may play an important role in the 
discussion of the presence of a non-zero subgap DoS resulting from mesoscopic 
fluctuations.\cite{Os} Considering a finite spin-flip scattering rate 
$\Gamma_{\rm{sf}}$, we find that the second branch of the solution no longer 
diverges at zero energy. If we increase the spin-flip rate up to a critical 
value the two zero energy solutions merge at $\Gamma_{\rm{sf}}^{c}$. 
The critical value can therefore be determined taking the limit $\beta^{\,1}\rightarrow0$ 
of the integral in the l.h.s. of equation (\ref{inte}) at zero energy.

\begin{figure}
    \includegraphics{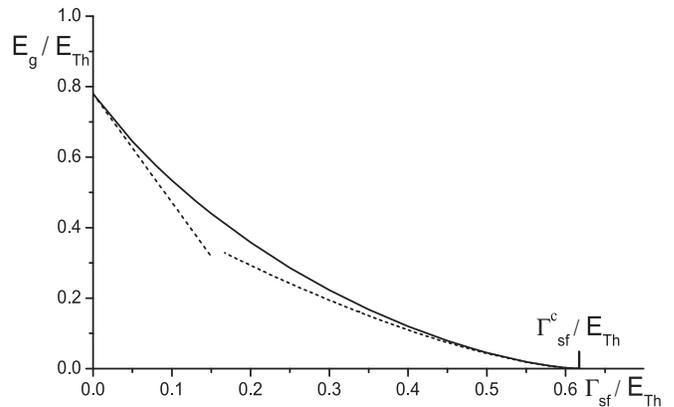}
    \caption{\label{num} Numerical gap curve (solid line): comparison with asymptotic expressions (dashed lines).}
\end{figure}
The complete dependence of the minigap on the spin-flip rate (Fig.\ \ref{num}) can be obtained by a simple numerical integration of equation (\ref{inte}). But it is possible to derive the asymptotic form of the gap curve for $\Gamma_{\rm{sf}}\rightarrow0$ and for $\Gamma_{\rm{sf}}\rightarrow\Gamma_{\rm{sf}}^{c}$.

In the limit of a small spin-flip rate, we can expand the integrand in  (\ref{inte}) in the small parameter $\frac{\Gamma_{\rm{sf}}}{\epsilon}$. Denoting $\hat{\beta}^{\,1}\approx1.421$ the value of the proximity angle corresponding to the zero spin-flip value of the minigap $E_{g}^{0}$, we find the resulting correction
\begin{equation}\label{f8}
E_{g}\approx\left(E_{g}^{0}-C_{1}\Gamma_{\rm{sf}}\right),
\end{equation}
where the coefficient $C_{1}$ is given by
\begin{equation}\label{f9}
C_{1}=\frac{\int_{0}^{\hat{\beta}^{\,1}}\frac{\left(\sinh\hat{\beta}^{\,1}+\sinh\beta\right)}{\left(\sinh\hat{\beta}^{\,1}-\sinh\beta\right)^{1/2}}d\beta}{\int_{0}^{\hat{\beta}^{\,1}}\frac{d\beta}{\left(\sinh\hat{\beta}^{\,1}-\sinh\beta\right)^{1/2}}}\approx3.09.
\end{equation}
The minigap decreases linearly with increasing spin-flip scattering rate.
From the magnitude of $C_{1}$, we can see that even a small spin-flip rate strongly affects $E_{g}$.

To obtain the analytic behavior close to $\Gamma_{\rm{sf}}^{c}$ is more tricky as all $\delta\Gamma_{\rm{sf}}=\Gamma_{\rm{sf}}^{c}-\Gamma_{\rm{sf}}$, $\beta$ and  $\frac{\epsilon}{\Gamma_{\rm{sf}}}$ are small. 
Following the procedure detailed in the Appendix we obtain the asymptotic dependence of the minigap near the closing point

\begin{equation}\label{a18}
E_{g}\approx2\left[\frac{2\,\delta\Gamma_{\rm{sf}}}{3}\right]^{3/2}.
\end{equation}

In Fig.\ \ref{num}, we compare the asymptotics (\ref{f8}) and (\ref{a18}) with the numerical gap curve. Dimensions are reintroduced in the graphs for clarity.

For $\Gamma_{\rm{sf}}>\Gamma_{\rm{sf}}^{c}$ the DoS in the N-region is finite at any energy. In this domain the Usadel equation (\ref{usad}) at zero energy has a solution with $\theta$ real. Applying again the procedure of the first integral (\ref{first}), but this time for a real $\theta$, we get
\begin{equation}\label{f12} 
\int_{\frac{\pi}{2}}^{\theta(x)}\frac{d\theta}{\sqrt{\cos^2\theta^{\,1}-\cos^2\theta}}=-2\sqrt{\Gamma_{\rm{sf}}}\;x.
\end{equation}
Inverting this elliptic integral, we can find a complete zero energy solution for equation (\ref{usad}). This solution involves the function $\textrm{dn}(u,k)$, which is one of the Jacobi elliptic functions, defined as inversions of the canonical forms of elliptic integrals (we follow the notations used in Refs.~\onlinecite{Law,Si})

\begin{equation}\label{ell4}
\theta(x)=\arcsin\left[\frac{\sin\theta^{\,1}}{\textrm{dn}\left(2\sqrt{\Gamma_{\rm{sf}}}(x-1),\cos\theta^{\,1}\right)}\right],
\end{equation}
where $\theta^{\,1}$ can be determined imposing the rigid boundary condition (\ref{bc1}) at $x=0$
\begin{equation}
\sin\theta^{\,1}=\textrm{dn} \left [ -2 \sqrt{\Gamma_{\rm{sf}}},\cos\theta^{\,1}\right].
\end{equation}

\begin{figure}
    \includegraphics[width=8.6cm]{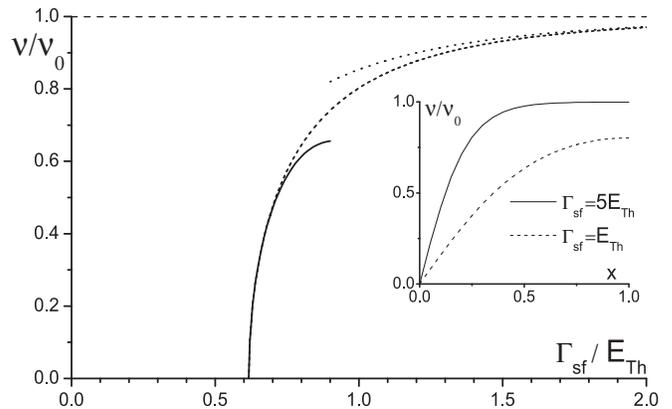}
    \caption{\label{num2} Dependence of the zero energy local DoS, at the open boundary, on the spin-flip scattering rate: asymptotic expression near $\Gamma_{\rm{sf}}^{c}$ (solid line), complete elliptic solution (dashed line) saturating to the normal state bulk value $\nu_{0}$ and its asymptotics at large $\Gamma_{sf}$ (dots). The inset shows the dependence on position of the DoS for $\Gamma_{\rm{sf}}=E_{\rm{Th}}$ and $\Gamma_{\rm{sf}}=5\,E_{\rm{Th}}$.}
\end{figure}

In the inset of Fig.\ \ref{num2}, we use relation (\ref{ell4}) to represent the dependence on position of the DoS in the N-region for two different spin-flip rates. As expected the density of states increases when we move away from the interface. For large spin-flip scattering rates, the elliptic solution (\ref{ell4}) saturates to the normal state bulk value $\theta(x)=0$, everywhere except in a thin domain close to the S-N interface (rigid boundary at $x=0$). 

Near the critical spin-flip rate, we found a square root dependence of the zero energy local DoS on $\Gamma_{\rm{sf}}$. Expanding the integrand in (\ref{f12}) in the small $\cos\theta^{\,1}$, we obtain that for $\Gamma_{\rm{sf}}\rightarrow\Gamma_{\rm{sf}}^{c}$ the density of states at the interface with vacuum is given by
\begin{equation}\label{ell14}
\frac{\nu(\epsilon{=}0,x{=}1)}{\nu_{0}}=\sqrt{2}\left[\frac{\delta\Gamma_{\rm{sf}}}{\Gamma_{\rm{sf}}^{c}}\right]^{1/2}-\frac{11}{8\sqrt{2}}\left[\frac{\delta\Gamma_{\rm{sf}}}{\Gamma_{\rm{sf}}^{c}}\right]^{3/2}+\ldots
\end{equation}

In the limit of large spin-flip rates, when the density of states approaches the normal state one, the expansion\cite{math} of the elliptic integral in (\ref{f12}) near $\cos\theta^{\,1}=1$ leads to an asymptotic expression for the DoS at the interface with vacuum
\begin{equation}\label{ell15}
\frac{\nu(\epsilon{=}0,x{=}1)}{\nu_{0}}=1-8\,e^{-4\sqrt{\Gamma_{\rm{sf}}}}+\ldots
\end{equation}
In Fig.\ \ref{num2}, we compare the expressions (\ref{ell14},\ref{ell15}) for the DoS at the open boundary with values obtained using the complete zero energy elliptic solution (\ref{ell4}).\\

In summary, 
we have studied the effect of spin-flip scattering on the minigap and 
zero energy density of states of a normal slab connected to a 
superconducting electrode. We have obtained analytic expressions for the 
dependence of the minigap on the scattering rate both for small spin-flip 
scattering rates and close to the critical value at which the gap closes. This critical value is controlled by the Thouless energy. For values larger
 than $\Gamma_{\rm{sf}}^{c}$ we find finite DoS at zero energy with a 
square root dependence on $\Gamma_{\rm{sf}}$. These results are valid for 
long S-N or S-N-S junctions with 
transparent interfaces, at small energies. The results on the minigap are 
not restricted to the case in which there is no spin-flip scattering in 
the superconducting regions. The only requirement is that the Thouless 
energy of the normal part is small compared to the gap in the superconducting one.
These results allow easy comparison with experimental measurements.

We thank M. Sigrist for useful discussions and acknowledge financial support from NCCR MaNEP of the Swiss Nationalfonds and the Spanish Science and Education Ministry through Ram\'on y Cajal contract.
\appendix*
\section{Gap curve close to $\Gamma_{\rm{sf}}^{c}$}
 
In this appendix, we derive the asymptotic form of the minigap curve close to $\Gamma_{\rm{sf}}^{c}$. We have seen previously that the minigap $E_{g}$ is the largest energy compatible with equation (\ref{inte}). Introducing
\begin{eqnarray*}
z_{\phantom{1}}&=&\sinh\beta^{\phantom{\,1}}+\alpha\\
z_{1}&=&\sinh\beta^{\,1}+\alpha\\[0.2cm]
\textrm{with}\quad\alpha&:=&\frac{\epsilon}{2\Gamma_{\rm{sf}}}\label{aa},
\end{eqnarray*}
we can rewrite equation (\ref{inte}) in the form
\begin{equation}\label{a4} 
2\sqrt{\Gamma_{\rm{sf}}}=\int_{\alpha}^{z_{1}}\frac{1}{\sqrt{z_{1}^2-z^2}}\cdot\frac{1}{\sqrt{1+(z-\alpha)^2}}dz.
\end{equation}
The integral in the r.h.s. is a function of $z_{1}$ and $\alpha$, which we denote $Y(z_{1},\alpha)$. To determine the minigap, we will find the maximum value of $Y$ over $z_{1}$ for a given value of the parameter $\alpha$. 

The critical spin-flip scattering rate (\ref{ss}) can be obtained setting $\alpha=0$ and maximizing (\ref{a4}). It turns out that $Y(z_{1},\alpha)|_{\alpha=0}$ is largest for $z_{1}=0$. The next step is to go to finite energies and expand $Y$ in $\alpha$. We write
\begin{equation}\label{a10}
Y(z_{1},\alpha)\approx{Y}(z_{1},\alpha)|_{\alpha=0}+\frac{\partial{Y}(z_{1},\alpha)}{\partial\alpha}|_{\alpha=0}\,\alpha,
\end{equation}
where
\begin{equation}\label{a11}
\frac{\partial{Y}(z_{1},\alpha)}{\partial\alpha}|_{\alpha=0}=-\frac{1}{z_{1}}+\frac{z_{1}}{1+z_{1}^{2}}.
\end{equation}
For a small $\alpha$, the maximum of $Y$ is expected to be close to the zero energy value $z_{1}=0$ and the second term $\sim{O}(z_{1})$ can be neglected. The first term in the r.h.s. can be expanded
\begin{equation}\label{a14}
Y(z_{1},\alpha)|_{\alpha=0}\approx\int_{0}^{1}\frac{1}{\sqrt{1-s^2}}\left[1-\frac{(z_{1}s)^2}{2}\right]ds.
\end{equation}
where $s=\frac{z}{z_{1}}$.
Taking the derivative of $Y$ over $z_{1}$, using the expansion (\ref{a10}),
we find that the maximum of Y is obtained for
\begin{equation}\label{a16}
\hat{z}_{1}=\left(\frac{\alpha}{\sqrt{\Gamma_{\rm{sf}}^{c}}}\right)^{1/3}.
\end{equation}
substituting back this result in (\ref{a10}) and using the definitions of $\alpha=\frac{\epsilon}{2\Gamma_{\rm{sf}}}$ and $Y=2\sqrt{\Gamma_{\rm{sf}}}$ we obtain an expression for $E_{g}(\Gamma_{\rm{sf}})$. Finally we write $\Gamma_{\rm{sf}}=\Gamma_{\rm{sf}}^{c}-\delta\Gamma_{\rm{sf}}$ and expand in the small $\delta\Gamma_{\rm{sf}}$ to get the asymptotic dependence (\ref{a18}).

\bibliography{spin_flip}

\end{document}